\documentstyle[12pt]{article}
\newcommand{\br}{\begin{array}}
\newcommand{\er}{\end{array}}
\newcommand{\beq}{\begin{equation}}
\newcommand{\eeq}{\end{equation}}

\newcommand{\N}{{\cal N}}

\newcommand{\ra}{\rightarrow}

\newcommand{\un}{\underline}

\def\ba{\begin{eqnarray}}
\def\ea{\end{eqnarray}}
\newcommand{\be}{\begin{equation}}
\newcommand{\ee}{\end{equation}}
\newcommand{\bea}{\begin{eqnarray}}
\newcommand{\eea}{\end{eqnarray}}
\newcommand{\we}{\wedge}

\makeatletter
\@addtoreset{equation}{section}
\makeatother

\def\appendix#1{
  \setcounter{section}{1}
  \setcounter{equation}{0}
  \renewcommand{\thesection}{\Alph{section}}
  \section*{Appendix \thesection\protect\indent \parbox[t]{11.715cm} {} }
  \addcontentsline{toc}{section}{Appendix \thesection\ \ \ }}

\begin{document}
\begin{flushright}
YITP-SB-00-90\\hep-th/0101064
\end{flushright}

\begin{center}
{\LARGE The polarization of M5 branes and little string theories with reduced 
supersymmetry }
\vskip1truecm

{\large\bf Iosif Bena\\ 
{\it University of California, Santa Barbara CA 93106}\\
Diana Vaman\\
{\it C.N.Yang Institute for Theoretical Physics,\\ S.U.N.Y. Stony Brook, 
NY 11794-3840, 
USA }}\footnote{ 
E-mail addresses: iosif@physics.ucsb.edu\protect\\
dvaman@insti.physics.sunysb.edu
} 
\vskip2truecm
\end{center}
\abstract{ We 
 construct an M-theory dual of a 6 dimensional little string theory 
 with reduced supersymmetry, along the lines of Polchinski and Strassler. 
 We find that upon perturbing the (2,0)
  theory with an R-current, the M5 branes polarize into a wrapped 
  Kaluza Klein monopole, whose isometry direction is along the R
  current. We investigate the properties of this theory.}

\newpage
$${}$$
\section{Introduction}

$${}$$

A while ago Polchinski and Strassler \cite{ps} found a string theory 
dual of a four dimensional SYM theory with reduced supersymmetry. They
 perturbed the $\N=4$ theory by giving mass to the hypermultiplets. 
 In the dual bulk theory this corresponded to turning on transverse RR and NS 
 3 form field strengths which polarized \cite{myers} the D3 branes into 
 D5 or NS5 branes. This extension of the original AdS/CFT duality  
 \cite{ads} had implications in both directions. It allowed for a
  detailed string theory interpretation of domain walls, baryon vertices, 
confinement,
  etc., and it illustrated a way in which string theory can cure some 
  singularities in supergravity. 

Since then, papers have appeared generalizing the PS construction 
to theories in other dimensions (M2 branes \cite{ib1}, D2 branes 
\cite{ib2,ib3}), as well as to 4 dimensional theories on orbifolds 
\cite{aharony}, with temperature \cite{freedman}, even less susy 
\cite{zamora,kinar}, etc.

One of the possible extensions of \cite{ps} left unexplored has been 
to the $AdS_7 \times S^4$ geometry. In the same line of thought from 
\cite{ps,ib1,ib2} one would expect the dual of a little string theory 
with reduced supersymmetry to be given in terms of an M theory bulk 
containing polarized M5 branes. It is the purpose of this paper to explore 
this possibility.

Since the degrees of freedom of a large number of M5 branes are not 
known, giving a matrix description to this polarization (like in the 
D-brane $\rightarrow$ D-brane polarization) is not possible. The only way to 
describe this polarization is thus to find the action of the object in 
which the M5 branes are polarized, and to show that there exists a 
configuration with nonzero radius which is energetically stable.

In order to understand which operators one needs to turn on to obtain
polarized  M5-branes, and find what they might polarize into, 
it is worth thinking about
the polarization of a large number of D4 branes into D6 branes. These 
configurations exist and are understood very well (they are T-dual to the 
D3 $\ra$ D5 system discussed in \cite{ps}). They are produced by turning on a 
fermion scalar bilinear on the worldvolume of the D4 branes, which corresponds 
to a combination of RR 1 form and NS-NS 2 form on the transverse space 
\footnote{We 
should note, however, 
that the naive expectation that the D4-D6 polarization configuration is still 
supersymmetric 
fails.}.

One expects upon lifting this polarized configuration to M theory to obtain a 
configuration in which M5 branes polarize into Kaluza-Klein monopoles (KKM). 
One of 
the directions of the M5 brane is special in that it is along the isometry 
direction of the monopole. The fermion scalar bilinear on the D4 brane can 
come from the 
dimensional reduction of a fermion scalar or a fermion vector 
bilinear oriented in the M-theory
direction. The bulk  RR 1 form and NS-NS 2 form come from a graviton with one 
leg
on the 11'th direction (the graviphoton) and respectively from a 3 form 
potential
with one leg on the 11'th direction. Since these bulk fields transform as 
vectors
under the Lorentz SO(5,1) group, we expect them to correspond to a 
fermion bilinear which is a vector oriented in the 11'th direction.
This same polarization configuration can be obtained by T-dualizing the D3 
$\ra$
NS5 polarization of \cite{ps} and obtaining a D4 brane polarized into a IIA KK 
monopole. The D4 $\ra$ IIA KKM configuration lifts to the same configuration 
as
above, with the M-theory direction identified differently.

One might wonder whether there are other possible polarizations for M5 branes. 
It appears possible to polarize it into a codimension 4 higher object, 
namely an M9-brane. However, since we are interested in M-theory 
configurations only, and the M9 is a brane solution of {\it massive} 
11d supergravity (sugra), we shall not explore this possibility. 

Nevertheless, it does not appear possible to polarize the M5's into a KKM 
which
does not have its isometry direction along the brane, simply because the 
original
M5 branes do not have that isometry.

There are however a few problems which an astute reader might already be 
concerned with. The first one has to do with the tension of the KK monopole. 
As we shall discuss, the tension of the KKM is proportional to the square 
of the length of 
the 11'th dimension. Thus, as we open up the 11'th dimension the monopole 
becomes
more and more massive. If for example in IIA we had $N$ D4 branes polarizing 
into
one D6 brane, as we open the 11'the dimension the ratio of the charges  
changes, and at one point the KKM in which the M5 branes polarize will be 
heavier
than the 5-branes, in which case it will not be treatable as a probe. We will 
address these
problems in due time.

While this paper was being completed, \cite{loz2} appeared, 
in which the polarization of M5 branes into KKM's was also suggested. 

\section{Perturbing the $AdS_7 \times S_4$ background}

\subsection{The sugra set-up}
$${}$$

The 11 dimensional supergravity bosonic Lagrangian is
\be
{\cal L}=R*1-\frac{1}{2}F_4\we * F_4+\frac{1}{6}F_4\we F_4\we
C_3 \label{L}
\ee
where $F_4=dC_3$, and $*$ is the 11-dimensional Hodge star. 
One can define the 6 form potential $C_6$ as the Poincar\'e dual of $C_3$
\be 
F_7 \equiv dC_6 =*F_4+\frac{1}{2} C_3\we F_4\label{f7}
\ee

The M5 brane is a soliton solution of 11d sugra
\bea
(ds^2)^{\bf 0}&=&Z^{-1/3}\eta_{\mu\nu}dx^\mu dx^\nu+Z^{-1/3}dx_{11}^2+
Z^{2/3}dx^i dx^j\delta_{ij}\label{metbac}\\
F_7^{\bf 0}&=&\partial_i Z^{-1} dx^0\we\dots\we dx^4 
\we dx^{11} \we dx^i \label{f7bac}
\eea
where the brane is aligned along $\mu=0,1,2,3,4,11$, the transverse 
coordinates are  $i=5\dots 9$, and $Z$ is the harmonic function on the 
transverse space.

It is well known that the background geometry of a large number
of coinciding M5 branes is $AdS_7\times S_4$, and that
11d sugra compactified on $AdS_7\times S_4$ gives a dual
description to the 6d (2,0) CFT theory living on the 
worldvolume of the M5 branes. 
For $N$ coincident M5 branes the harmonic function is
\be
Z=\frac{R^3}{r^3},\;\;\;\;\;\; r^2=x^i x^i,\ \ \ \ \
R^3 \equiv \pi N l_p^3 
\ee
The $AdS_7\times S_4$ geometry becomes transparent if we redefine 
$r=4R/z_0^2$. The $AdS_7$ radius is thus twice
the $S_4$ radius. For later need, we give the Poincar\'e dual of (\ref{f7bac})
\be
F_4^{\bf 0}= \frac{1}{4!}\epsilon_{ijklm}\partial_m Z dx^i\we
dx^j\we dx^k\we dx^l\label{f4bac}
\ee
where $\epsilon_{ijklm}$ is the flat antisymmetric tensor restricted 
to the transverse coordinates.

We want to perturb this background with a non-normalizable mode of the field
which is dual to a fermion mass operator in the CFT.
The fermions of (2,0) CFT transform in the 
$\underline{4}$ spinor representation of the R-symmetry group
$SO(5)$. Thus a fermion bilinear which is a $\underline{10}$ 
of $SO(5)$ corresponds to a 2-form (or by Poincar\'e duality a 3-form) 
field strength perturbation in the space transverse to the branes.
Since 11d SUGRA has only a 4-form field strength and a graviton, the only 
way to obtain a transverse 3-form is to consider a 4 form with a leg along 
one of the longitudinal directions. Similarly the transverse 2-form is the 
field strength of the graviphoton.

We can see that only a fermion
bilinear which is a Lorentz vector is a chiral primary. Thus, to turn on
a fermion bilinear which is chiral we need to break Lorentz covariance 
and introduce a preferred direction, or equivalently, 
a Killing vector.  We shall see later that this breaking 
of Lorentz covariance is the key to the construction in which the M5
branes polarize into Kaluza-Klein monopoles.

The background geometry admits a Killing vector
\be
D_{(\Lambda}^{\bf 0} k_{\Pi )}=0, \;\; \Lambda=0\dots 9,11
\ee
with $k_{\Lambda}=Z^{-1/3}\delta_{\Lambda 11}$. We note that the
contravariant Killing vector is simply $k^\Lambda=
\delta^{\Lambda 11}$. From now on we will use $^{\bf 0}$ to denote 
background fields and ${}^{\bf 1}$ to denote perturbed fields.

We want to turn on the nonnormalizable modes of the perturbations 
$\delta g_{i 11}\equiv g_{i 11}^{\bf 1} \equiv h_{11 i}$
and $\delta C_{11 ij}^{\bf 0}\equiv C^{\bf 1}_{11 ij}$
 which only depend on the transverse coordinates.
The metric perturbation is related to the 3-form perturbation by 
\bea
\delta R_{11 i}&=&\delta\left\{\frac{-1}{12}(F_{11 ...} 
{F_{i}}^{...}-\frac{1}{12}g_{11 i} F_{....}F^{....})\right\}.
\label{deltar}
\eea
The variation of the Ricci tensor
\be
\delta R_{\Lambda\Pi}=\frac{1}{2}\left(h^{\Omega}_{\;\Lambda;
\Pi\Omega}+h^{\Omega}_{\;\Pi;\Lambda\Omega}-h_{;\Lambda\Pi}-
\Box h_{\Lambda\Pi}\right)
\ee
in the background (\ref{metbac}) is 
\be
\delta R_{11 i}=\frac{1}{2}\partial_{j}\left(Z^{-1}\partial_j(
h_{11 i} Z^{1/3})-Z^{-1}\partial_{i}(h_{11 j} Z^{1/3})\right)-
\frac{1}{24}h_{11 i}\partial_m Z\partial_m Z Z^{-8/3}
\ee
Combining this with (\ref{deltar}) we obtain   
\be
d[Z^{-1} (*_5 d(h Z^{1/3})+ G_3)]=0\label{har1}
\ee
where $*_5$ denotes the flat Hodge dual in the transverse space, 
$G_3\equiv (1/3!)F_{11 ijk}^{\bf 1} dx^i\we dx^j \we dx^k$, and 
 $h\equiv h_{11 i}\we dx^i$.

Similarly, from the first order variation of the $F_4$ field equation:
\be
d *^{\bf 0} F_4^{\bf 1} + d *^{\bf 1} F_4^{\bf 0}=0
\ee
we can obtain after a few straightforward steps:
\be
d[Z^{-1} (d(h Z^{1/3}) +*_5 G_3)]=0. \label{har2}
\ee

In order to find the exact form of the bulk 2 and 3-forms corresponding to 
the  
fermion bilinear we need to relate them to transverse tensors, and to 
solve  (\ref{har1}) and (\ref{har2}).
\subsection{Fermion bilinears and transverse tensors}

$${}$$

The fermion R-current can be related to a tensor on the transverse space by 
analyzing their properties 
under the action of the R-symmetry group. 
We pair the 4 worldvolume fermions 
and 4 of the transverse space coordinates into complex variables
\bea
&&z_1= x^5+i x^6\;\;\;\;\;\; z_2=x^8+i x^9\\
&&\Lambda_1=\lambda_1+i\lambda_3\;\;\;\;\; \Lambda_2=\lambda_2+i\lambda_4
\label{complexified}
\eea
and we notice that under an SO(5) rotation $z_i\rightarrow e^{i\phi^i}z_i$,
the fermions transform as
\bea
&&\Lambda_1\rightarrow e^{i(\phi^1-\phi^2)/2}\Lambda_1\\
&&\Lambda_2\rightarrow e^{i(\phi^1+\phi^2)/2}\Lambda_2
\eea
Thus, a diagonal fermion ``mass'' term behaves in the same way under SO(5)
rotations as
\be
T_2= -{\rm Re} [m_1 dz_1\we d\bar z_2 +m_2 dz_1\we dz_2]\label{t2}
\ee
To make $T_2$ a Lorentz vector on the worldvolume all we have to do is 
tensor it with $\we dx^{11}$.
We can also define on the transverse space the Hodge dual to $T_2$
\be
T_3=*_5 T_2= {\rm Re} [m_1 dz_1\we d\bar z_2 +m_2 dz_1\we dz_2]\we dx^7. 
\label{t3}
 \ee
Note that if we set either fermion mass to zero, there is a remaining
global $U(1)$ symmetry. 

\subsection{The first order perturbation}

$${}$$

A general bulk 3-tensor with the SO(5) transformation properties of a fermion 
bilinear 
will be a combination of $T_3$ and $V_3$, multiplied by a power of $r$, where
$V_3$ is given in the Appendix.
Thus, we can express the bulk 3-tensor $H_3$ and 2-tensor 
$F_2 \equiv d(h Z^{1/3})$ as 
\bea
F_2&=&\left({\frac{r}R}\right)^p(aV_2+bT_2)\nonumber\\
G_3&=&\left({\frac rR}\right)^p(AV_3+BT_3)\label{decomp}
\eea
where $T_2$ and $T_3$ were found above, and $V_2$, $V_3$ together with some 
useful identities are 
given in the Appendix. Using the Bianchi identities for the field strengths 
$F$ and $G$ we get the 
constraints
\bea
pb&=&2a,\nonumber\\
pB&=&3A\label{rel}.
\eea
Substituting (\ref{decomp}), (\ref{rel}) into the harmonic
form eq. (\ref{har1}), (\ref{har2}) we obtain the system:
\bea
(p+3)(b+pb/2+B)-3(pB/3-pb/2)&=&0,\nonumber\\
(p+3)(b+B+pB/3)-2(pB/3-pb/2)&=&0,
\eea
which admits four real solutions: $(p=0, b/B=-1), (p=-8, b/B=-1);
(p=-5, b/B=2/3), (p=-3, b/B=2/3)$.

To interpret these modes, we need to know their conformal dimensions.
To translate  $G_3=(1/3!)\partial_i C_{11 jk}^{\bf 1} dx^i\we dx^j \we dx^k$ 
to an inertial frame, we multiply (\ref{decomp}) with 
$\sqrt{G^{11,11}G^{ii}G^{jj}G^{kk}} = Z^{-5/6} \propto r^{5/2}$.

Remembering that in AdS coordinates $z_0 \sim r^{-1/2}$, 
we find that the modes of $G_3$ corresponding to $p=-3$ and $p=-5$ 
behave near the boundary as $z_0$ and $z_0^5$ respectively. 
Thus, they are the non-normalizable and respectively normalizable solutions 
corresponding to an operator of conformal dimension $\Delta=5$, which is 
our fermion bilinear. 
By matching the $R$-symmetry representation and conformal dimension
with the table of CFT operators given in \cite{lr}, we find that the CFT
operator is indeed a Lorentz vector, bilinear in fermions - the $R$-symmetry 
current. 
In previous situations \cite{ps,ib1, ib2}, the 
CFT operator corresponding to perturbations in the bulk matching the 
$R$-symmetry representation of a fermion bilinear was always a Lorentz 
scalar. In our case, the only bilinear in fermions which is a chiral primary 
operator is 
the $R$-current. The corresponding bulk  perturbations 
are vectors under the $SO(5,1)$ Lorentz group, as anticipated.
The non-normalizable mode $p=-3$ corresponds to the fermion  
``mass'' perturbation, 
while the normalizable mode $p=-5$ corresponds to the vev $<\psi\psi>$ 
\cite{bdhm-bklt}.
Thus an R-current
corresponds to the bulk perturbation:
\be
Z^{-1} [*_5 d(h Z^{1/3}) +G_3)] = \xi T_3\label{comb}
\ee
where $\xi$ is a numerical constant which relates the physical mass of the 
fermions to the coefficient of the bulk mode. From now on we will absorb $\xi$
in the fermion mass, and thus set it to 1 in the bulk 
\footnote{Let us briefly discuss how one can relate the fermion ``mass'' term 
of the theory living on the boundary
of $AdS_7$ to the mass parameter of the bulk perturbations (\ref{comb}, 
\ref{t2}, \ref{t3}).
We start with the action of the M5-brane in an arbitrary sugra background
\cite{claus}.
After fixing the $\kappa$-symmetry, half of the $\theta$ superspace 
coordinates are eliminated, while the other half become fermions on the
worldvolume. Thus, by specifying our perturbed $AdS_7$ background, we can
search for terms bilinear in fermions. For instance, a Lorentz vector
``mass'' term comes from the WZ term (specifically  from the pull-back of 
$C_6(X,\theta)$, where $E_M^{\underline A}$ is no longer zero 
by gauge choice). However, since the only parameter we introduced is the 
bulk mass (in units where the AdS radius is one), 
this implies that we have at most
a numerical constant relating the boundary mass to the bulk mass. }. 
We only wrote down the 
combination of modes which later will appear in the KKM Lagrangian. 
It is interesting to notice that this particular combination is a harmonic 
function (as one can easily see from (\ref{har1}) and (\ref{har2})). Therefore,
it is determined by its value at the boundary (\ref{comb}), and does not 
depend on the particular form of $Z$.

Similarly, the modes with  $p=-8$ and $p=0$ correspond to a boundary 
operator of conformal dimension  $\Delta =11$. We can identify this operator
to be  $ HH\psi\psi$, where $H$ is the selfdual 3-form of the 6d (2,0) 
tensor multiplet. 

One might wonder if our perturbation (\ref{comb}) does not become stronger 
than
the 
background is some regime. We will discuss this possibility in section 3.2.

\section{Polarizing the M5 branes into KK Monopoles}

$${}$$

We would like to investigate whether a large number of M5 branes can polarize 
into KK monopoles. If the self-interacting potential of a KKM wrapped on a 2 
sphere, with a very large M5 charge has a minimum when the 2-sphere radius is 
nonzero, polarization occurs. In order to find this self interacting potential 
we
first find the action for a test KKM with large M5 charge ($n$), in the 
potential
created by a very large number $N>>n$ of M5 branes. Anticipating, this test 
potential will be found to be independent of the positions of the $N$ M5 
branes. 
The full self interacting potential can be found by bringing from infinity 
test
KKM/M5 shells. Since the potential of these shells does not depend on the 
positions of the other M5 branes, the probe calculation with $n$ replaced by 
$N$ will give the full self-interacting potential.  

\subsection{The KKM action}

$${}$$

As we explained in the introduction, a Kaluza-Klein monopole in 11d 
supergravity is an object which has 7 longitudinal directions, 3 transverse, as
well as a special isometry direction. When one compactifies along this isometry
direction one obtains a D6 brane in IIA. If we express the D6 brane tension in 
terms of the radius of the 11'th direction ($R_{11}$) and the 11'th 
dimensional
Planck length $l_P$, we find that it is proportional to $R_{11}^2$:
\be
T_{D6}=\frac{1}{ g (2\pi)^6\alpha'^{7/2}}=\frac{R_{11}^2}{l_P^9 (2\pi)^6}
\label{alpha'}
\footnote{
Using $g=l_P^{-3/2} R_{11}^{3/2}$ and $\alpha'=l_P^{3} R_{11}^{-1}$.}.
\ee
 Therefore, the
KKM cannot be thought of as some 7 brane wrapped on the isometry direction. 
This direction is special, and gets special treatment in the action. 

The action for a Kaluza-Klein monopole (KKM) was given by \cite{bjo,bel}.
As expected, it depends on the  background geometry Killing vector $k^\Lambda$ 
corresponding to the isometry direction.

In the unperturbed  $AdS_7 \times S_4$ background this Killing vector was
found to be ${k^\Lambda}^{\bf 0}= \delta^{\Lambda 11}$. 
In the perturbed background, the variation $\delta k^{\Lambda}\equiv
{k^{\Lambda}}^{\bf 1}$ can be found by solving
\be
D_{(\Lambda}^{\bf 0} k^{\bf 1}_{\Pi )}=\frac{1}{2} (h_{11\Lambda;\Pi}+
h_{11 \Pi; \Lambda}-h_{\Lambda\Pi;11})\label{kilpert}
\ee
It may not come as a surprise that $k^{\Lambda}$ is still proportional with
$\delta^{\Lambda 11}$, as one can easily check. However, since this Killing
vector measures the length of the isometry direction at infinity, we 
normalize it as 
\be
k^\Lambda=\frac{R_{11}}{l_P} \delta^{\Lambda 11}
\ee

The KKM action is the sum of a Born-Infeld (BI) and a Wess-Zumino (WZ) piece.
The BI part (in an  11-dimensional sugra background) reads\footnote{
Identifying the reduced KKM action with the D6 action, not only allows us to
relate their tensions, but the BI actions as well. As a consequence, 
the factor $2 \pi \alpha'$ appears in the KKM action, with the understanding 
that
it is a parameter which depends on $l_P$ and $R_{11}$ as specified in 
(\ref{alpha'}).}
\be
S_{BI}=-\frac {1}{(2\pi)^6 l_P^7} \int d^7\xi 
k^2\sqrt{|\det (D_a X^\Lambda D_b X^\Pi g_{\Lambda\Pi}
+(2\pi \alpha') k^{-1} {\cal F}_{ab})|}
\ee
where the covariant derivative is defined with respect to a gauge field
$D_a X^\Lambda =\partial_a X^\Lambda +A_a k^\Lambda$ which is not an 
independent
field: $A_a=k^{-2}\partial_a X^{\Lambda} k_{\Lambda}$. 
The field strength of the
one-form field $b=b_a d\xi^a$ living on the worldvolume is 
\be
{\cal F}_{ab}= \partial_a b_b-\partial_b b_a + 1/(2\pi \alpha')
\partial_a X^\Lambda \partial_b X^\Pi k^{\Sigma} C_{\Sigma\Lambda\Pi}
\ee
The WZ piece is 
\be
S_{WZ}=\frac{2\pi\alpha '}{(2\pi)^6 l_P^{7}}
\int K_{7}
\ee
where the $K_7$ is the field strength of a non-propagating worldvolume 
6-form
\bea
K_7&=&d\omega_6+\frac{1}{2\pi\alpha'}(i_{k}N_8)+(i_k C_6)\we {\cal F}
\nonumber\\
&&-\frac{2 \pi \alpha'}{(3!)^2}DX^\Lambda\we DX^\Pi\we DX^\Sigma 
C_{\Lambda\Pi\Sigma}
\we (i_k C_3)\we (i_k C_3)\nonumber\\&&-\frac{1}{3!}
DX^\Lambda \we DX^\Pi \we DX^\Sigma C_{\Lambda\Pi\Sigma}\we (i_k C_3)
\we db\nonumber\\&& -\frac{1}{2\cdot 3!}(2\pi\alpha') DX^\Lambda\we
DX^\Pi \we DX^\Sigma 
C_{\Lambda\Pi\Sigma}\we db\we db\nonumber\\
&&-\frac{1}{3!}(2\pi\alpha')^2 A\we db\we db\we db\label{k7}
\eea
where, following \cite{bel} we wrote explicitly  the pull-backs only when the 
covariant derivatives where involved. The 8-form $N_8$ is by definition
the Poincar\'e dual of the Killing 1-form.

The action of the KKM is invariant under certain gauge transformations.
We list only the ones which are relevant for this paper:
\bea
&&\delta C_3=d\chi_2\\
&&\delta i_k C_6=d i_k\chi_5-\frac{1}{2}d i_k \chi_2\we C_3+ 
\frac{1}{2}d \chi_2 \we i_k C_3 \\
&&\delta i_k N_8= di_k\chi_5\we (i_k C_3)+\frac{1}{3} d\chi_2\we i_k C_3
\we i_k C_3-\frac{1}{3} d i_k \chi_2\we C_3 \we i_k C_3\label{n8}
\eea
From (\ref{n8}) we infer a gauge invariant definition of the field strength
of $i_k N_8$:
\be
d i_k N_8 = i_k d*k - i_k C_6\we i_k F_4-\frac{1}{6} C_3\we i_k C_3\we i_k F_4
-\frac{1}{6} i_k C_3 \we i_k C_3\we F_4\label{N_8}
\ee
We can convince ourselves that by varying (\ref{N_8}) we get the same
contributions we obtain by taking an exterior derivative  
in (\ref{n8}).

\subsection{The KKM probe}

$${}$$

We begin by testing the perturbed background geometry
with a KKM probe. The monopole is wrapped on a 2-sphere in the (578) plane 
(anticipating, this will be one of the planes where polarization is most 
likely to
occur) 
has 5 flat directions along $x^0,\dots ,x^4$, and has an M5 charge $n \ll N$. 
The potential for more general 
orientations of the two-sphere will be considered in the following sections.
The isometry direction is 
$k^\Lambda = (R_{11}/l_P)\delta^{\Lambda 11}$. As discussed in the previous 
subsection 
this is a Killing vector of the perturbed $AdS_7\times S_4$ geometry. 

As one can see from the WZ term (\ref{k7}), the monopole will have M5 charge 
$n$
if  
we turn on its worldvolume 2-form field strength $F_2$:
\be
 \frac{1}{(2\pi)^6 l_P^7}\frac{R_{11}}{l_P}
\int_{S_2} (2\pi \alpha') F_2 =(2\pi R_{11}) T_{M5}\ n =
\frac{2\pi R_{11}}{(2\pi)^5 l_P^6} n,
\ee
which gives 
\be
(2\pi\alpha')F_2=(2\pi\alpha')F_{\theta\phi}d\theta\we d\phi=\frac{1}{2} 
{2n\pi l_P^2}\sin\theta d\theta \we d \phi
\ee

We work in the regime in which the M5 contribution to the BI action of the 
probe
monopole is dominant. Since we know the form of $F_2$, we can rewrite the BI 
action as
\be
-\frac{1}{(2\pi)^6 l_P^7} 
\int_{S_2} d\theta d\phi\  k^2 \sqrt{\det G_{||}} \sqrt{\det G_{ab} + \det 
\left(\frac{(2\pi\alpha'){\cal F}_{ab}}{k}\right)}
\ee
where $G_{||}$ denotes the induced metric on the 5 space-time directions
common to the KKM and the M5 brane, and $a,b$ are coordinates on the  
2-sphere.
We have assumed the  integration over $dx^0\dots dx^4 $ to be implicit, and we
will assume this throughout this section.
Using the induced metric $G_{ab}=\{G_{\theta\theta}=Z^{2/3} r^2 , 
G_{\phi\phi}=Z^{2/3} r^2 \sin^2\theta\}$, we find the condition that the M5 
contribution to the
BI action dominates to be
\be
\det\left(\frac{
(2\pi\alpha '){\cal F}_{ab}}{k}\right)=\frac{(2\pi\alpha ')^2}{2k^2}
\det(G_{ab}){\cal F}_{ab} {\cal F}^{ab} \gg \det G_{ab},
\ee
or
\be
\frac{(2\pi)^2 l_P^6 n^2 Z^{-4/3} r^{-4} Z^{1/3} }{4 R_{11}^2} \gg 1 
\footnote{Note this condition is 
not independent of $r$ as one might expect in a conformal theory. The source 
of
nonconformality is the 
dependence of the effective size of the 11'th dimension (which is also the KKM 
mass) on $r$.}.
\label{cond}
\ee
Under this assumption we can Taylor expand the BI action 
\be
-\frac{1}{(2\pi)^6 l_P^7} 
\int_{S_2} d\theta d\phi k^2\sqrt{\det G_{||}}
\sqrt{\det 
\left(\frac{(2\pi\alpha'){\cal F}_{ab}}{k}\right)}\left[
1+\frac{k^2\det G_{ab}}{2(2\pi\alpha')^2\det{\cal F}_{ab}}\right]\label{bi}
\ee
and find the leading contribution 
\be
-\frac{R_{11}}{2(2\pi)^5 l_P^6}\int_{S_2}d\theta d\phi
 Z^{-1} n \sin\theta=-\frac{R_{11}}{(2\pi)^4 l_P^6} Z^{-1} n \label{bil}
\ee
and the subleading one
\bea
-\frac{R_{11}^3}{(2\pi)^7l_P^{12}}\int_{S_2} d\theta d\phi
 Z^{-1/2} Z^{-5/6}\frac{ Z^{4/3}r^4\sin\theta}{n}
= -\frac{2 R_{11}^3}{(2\pi)^6 l_P^{12}} 
\frac{r^4}{n}.
\eea
The leading terms in the BI and WZ actions represent M5-M5 interaction, and 
cancel
as expected. More explicitly, the WZ leading contribution is
\be
\frac{1}{(2\pi)^6 l_P^7}\int i_k C_6^{\bf 0}\we (2\pi\alpha')F_2 =
\frac{1}{(2\pi)^5 l_P^6}\frac{R_{11}}{2}
\int_{S_2} d\theta d\phi Z^{-1}n \sin\theta,
\ee
and cancels against (\ref{bil}).

The subleading contribution to the WZ term comes from $\int i_k N_8$.
When the bulk graviphoton and three form perturbation is turned
on, several terms in (\ref{N_8}) ($i_k*^{\bf 0}dk^{\bf 1}+i_k*^{\bf 1}dk^{\bf 
0}
+i_k C_6^{\bf 0}\we i_k F_4^{\bf 1}$ ) become nonzero. We find
\be
d i_k N_8 =\frac{R_{11}^2}{l_P^2}\left[
Z^{-1} \left(*_5d(h Z^{1/3}) +G_3)\right)\right]\we dx^0\dots\we dx^4
\ee  
which has the solution $i_k N_8=  (1/3!)(R_{11}^2/l_P^2)
T_{mnp} x^m dx^n\we dx^p\we dx^0\dots\we dx^4$.
It is interesting to notice that $i_k N_8$ only depends on the combination
$Z^{-1} [*_5 d(h Z^{1/3}) +G_3)]$, which is independent of the distribution of
the M5 branes, as we discussed in section 2.2.

We can see from (\ref{t3}) that the WZ subleading contribution is the largest 
if the 2 sphere is in the (578) or the (679) planes (depending on the signs 
and
magnitudes of $m_1$ and $m_2$). We will assume from now on that both $m_1$ and 
$m_2$
are positive, and call their sum $m$.  Thus the test brane will have a local 
minima in both the (578) and (679) planes when  $m_2 \approx 0$, and only in 
the (578) for $m_2 \approx m_1$. 

Thus, the WZ contribution to the action is
\be
\frac{1}{(2\pi)^6 l_P^7}\int i_k N_8 =m\frac{{4\pi} R_{11}^2}{(2\pi)^6 l_P^9} 
r^3.
\ee
 Besides the BI and the WZ contributions found above there is yet another 
contribution 
of the same order which comes from the second order perturbation to the 
fields,
and which will be 
discussed in the next chapter. The polarization radius $r_0$ is of the same 
order
as the radius where  
$V_{BI}+V_{WZ}$ is minimized. Thus
\be
r_0\sim\frac{ nm l_P^3}{R_{11}}.\label{ro}
\ee  
Substituting in (\ref{cond}) we find that we can treat the KKM as a probe if
\be
n \gg R_{11} m N .\label{cond2}
\ee

We can also ask if the strength of the perturbation is anywhere larger than 
the energy density in the unperturbed AdS background 
\be
\frac{|{F_4}^{\bf 1}|^2}{|{F_4}^{\bf 0}|^2}\sim \frac{r^{-1} 
R m^2 }{R^{-2}}\sim \frac{m^2 l_P^3 N}{ r}\label{f4/f6}
\ee
The fraction (\ref{f4/f6}) grows as $r$ decreases towards $r_0$. Nevertheless, 
inside the shell we expect 
this ratio to be of the same order as at $r_0$.  By estimating (\ref{f4/f6}) 
at the polarization radius we can 
see that the perturbation is always small if (\ref{cond2}) holds.

We would like to examine more the physical implications of (\ref{cond2}), 
which is necessary for this 
polarization picture to be correct. Since in the final computation $n$ will be 
replaced by $N$, 
(\ref{cond2}) is equivalent to  $R_{11} m \ll 1$, which seems to imply that we 
are in fact describing only 5-dimensional physics. What we are describing is
in fact the perturbed (2,0) theory compactified on a circle of radius $R_{11}$.

We can estimate the effective radius of the 11'th dimension 
$R_{11}^{physical}=R_{11}Z^{-1/6}$ 
we find its lowest bulk value to be $N^{1/3} (m R_{11})^{1/2}l_p$ (except very 
near the KKM shell, when physics is described by the theory living on the 
KKM).
In the spirit of \cite{imsy}, this implies that for $N \gg (m R_{11})^{-3/2}$ 
(which
is the regime we will be considering), the 10 dimensional string coupling 
constant is large everywhere in the bulk, and thus 11 dimensional supergravity 
is
the appropriate perturbative description. For smaller energy scales the 
perturbative
description of our theory will be described by a IIA supergravity in the bulk, 
and for even smaller energy scales $R_{11} m \ll 1/N$ the super-Yang Mills 
theory
on the D4 branes.
We will also find that the bulk describes the 
properties of the objects living in this theory, which are the objects 
living on coincident M5 branes.

The fact that we can only have a polarization state in the regime 
$m R_{11} < 1$ seems puzzling if the theory we are dealing with is 
supersymmetric.
Indeed, one does not expect a vacuum to just go away, as we increase 
adiabatically
$m$ or $R_{11}$. Nevertheless, this is exactly what happens. The answer to this 
puzzle will be found in the next chapter - in order to polarize the branes one 
needs to break all supersymmetry.

\subsection{The full problem}

$${}$$

We use the results of the previous section to solve the full
problem of a self-interacting KKM with large M5 charge. The self-interaction 
potential can be found by bringing from infinity probe KKM's with M5 charge
in the background of polarized M5 branes. As we will immediately show,  the 
potential 
these probes feel is 
unaffected by the distribution of the M5 branes. Therefore, the total 
self-interaction potential will be the same as the potential of a single 
shell,
with $n$ replaced by $N$.

The geometry of a distribution of M5 will still
be given  by (\ref{f7bac}), but the warp factor $Z$ is a superposition of 
harmonic
functions. For the case of $N$ M5 branes uniformly distributed on a 2-sphere 
of
radius $r_0$ in the (578) hyperplane, the warp factor is
\bea
Z&=&\frac{1}{2}\int_0^\pi d\theta \frac{R^3\sin\theta}{(r_0^2+r_1^2+r_2^2-2r_0 
r_1
\cos\theta)^{3/2}}\nonumber\\
&=&\frac{R^3}{2r_0 r_1}\left(\frac{1}{\sqrt{(r_0-r_1)^2+r_2^2}}-\frac{1}{
\sqrt{(r_0 +r_1 )^2+r_2^2}}\right)\label{warp}
\eea
where $r_1$ is the radius in the (578) hyperplane and $r_2$ is the radius in 
the
(69)
plane.
For large $r_1, r_2$ we recover the $AdS_7$ warp factor. If the M5 branes are 
distributed
on several shells, then the total warp factor will be the sum of each shell 
$Z$
factor.

As we explained in Section 2, equations  (\ref{har1}) and (\ref{har2}) imply 
that $Z^{-1} [*_5 d(h Z^{1/3}) +G_3)]$ is a harmonic function, and therefore 
it is given by its value at infinity, regardless of 
the value of $Z$. This is exactly the combination that enters the 
WZ term of the 
KKM. Also, in the large 
$N$ limit, the BI term can be expanded as in (\ref{bi}), and the 
cancellation of  the leading WZ and BI 
takes place as before. The subleading terms are unchanged, and have no $Z$ 
dependence. We thus conclude, that for each M5 shell
the potential is the same as the probe potential.


The most general polarized configuration consistent with the symmetries is a 
KKM wrapped on a two-ellipsoid:
\be
x_7=\un{x_7}\cos \alpha, \ \
z_1=\un{z_1}\sin \alpha \cos \beta,\ \ 
z_2=\un{z_2}\sin \alpha \sin \beta,
\ee
where $z_1$ and $z_2$ are defined in (\ref{complexified}), $\alpha$ 
and $\beta$ parameterize the embedding, and $\un{x_7},\un{z_1}$ and $\un{z_2}$ 
give
the length and orientation the semiaxes.
It is a straightforward exercise to express the contribution of the BI and 
WZ potentials in terms of the semiaxes of the probe ellipsoid
\bea
V_{BI+WZ}&=&\frac{2 R_{11}^3}{3n (2\pi)^6 l_P^{12}}(|z_1|^2 x_7^2+ |z_2|^2 
x_7^2
+
|z_1|^2 |z_2|^2)\nonumber\\
&-&\frac{4\pi R_{11}^2}{(2\pi)^6 l_P^9} {\rm Re}(m_1 z_1  \bar z_2 x_7+
m_2 z_1  z_2 x_7),\label{v2}
\eea
where we have dropped the subscripts on the semiaxes to keep notation 
simple. From now and throughout this
chapter we will denote the semiaxes by $x_7,z_1$ and $z_2$ without an 
underline.

There is another term in this potential which comes from a second order
correction to the background. This term has the same relevance as the first 
two,
and is hard to compute explicitly (it has been done in \cite{freedman}). 
We have 
already discussed that by giving masses to either of the two fermions we break 
half of the
original amount of supersymmetry ((2,0) in 6 dimensions). Thus, 
when one of the fermion masses ($m_2$ 
in this case) is zero, we can use supersymmetry to complete the squares in 
(\ref{v2}). This gives the complete effective potential
\bea
V_{total} &=& \frac{2 R_{11}^3}{3n (2\pi)^6 l_P^{12}}\bigg[(|z_1|^2 x_7^2+ 
|z_2|^2 x_7^2
+
|z_1|^2 |z_2|^2)-\frac{6\pi n l_P^3}{R_{11}} {\rm Re}(m_1 z_1  \bar z_2 x_7)
\nonumber\\
&+&\left(\frac{3 \pi nl_P^3}{2 R_{11}}\right)^2 m_1^2(|z_1|^2 + |z_2|^2)\bigg],
\label{vt1}
\eea
is derived in the usual way from the superpotential $W$
\be
W=\sqrt{\frac{2 R_{11}^3}{3n(2\pi)^6 l_P^{12}}}\left[z_1 z_2 x_7 -
\frac{3n\pi l_P^3}{4 R_{11}}m_1
(z_1^2 +z_2^2)\right]
\label{w}
\ee 
The potential (\ref{vt1}) has only one minimum at $x_7=z_1=z_2=0$. We 
conclude that the polarization of M5 branes into a wrapped
KKM is not possible if we have any supersymmetry left. 

We remember that in 
\cite{ps}, one had to give mass to  at least 3 of the 4 complex Weyl fermions 
in order for polarization to occur. In our case (as well as in the  D4-D6 
case),
we only have two fermions, so one can either give mass to one (1/2 of the 
fermions) which preserves $\N=1$ supersymmetry but is not enough for 
polarization, or give mass to both fermions, break supersymmetry completely, 
and
polarize the branes
\footnote{Naively one might think that by T-dualizing the D3-D5 polarization
of \cite{ps} to a D4-D6 configuration we do not lose supersymmetry. We can see 
that this is not the case, both by investigating the theory on the brane 
(one needs to pair the D3 fermions two by two to obtain D4 brane 
fermions, and thus cannot consistently give mass to one fermion and a half), 
or by realizing that the bulk 
in \cite{ps} has no isometry along the T-duality direction.}.

There are two ways one can bypass this problem 
\footnote{Similar methods and a more thorough 
discussion can be found in \cite{ib2}.}. The first one is to consider an  
``almost supersymmetric'' 
situation, when the susy breaking mass $m_2$ is far smaller than $m_1$. One 
can
still use 
supersymmetry to complete the squares in (\ref{v2}), and find the potential to 
be
\bea
V &=& \frac{2 R_{11}^3}{3n (2\pi)^6 l_P^{12}}\bigg[(|z_1|^2 x_7^2+ 
|z_2|^2 x_7^2
+
|z_1|^2 |z_2|^2)-\frac{6 \pi n l_P^3}{R_{11}}\ {\rm Re}(m_1 z_1  
\bar z_2 x_7 + m_2 z_1  z_2 x_7)\nonumber\\
&+&\left(\frac{3 nl_P^3}{2 R_{11}}\right)^2 
(m_1^2|z_1|^2 + m_1^2 |z_2|^2 + 4 m_2^2 x_7^2)\bigg].
\label{vt2}
\eea

In this case, the M5 branes will polarize into a monopole wrapped on 
an ellipsoid. The new potential has a minimum at
\bea
z_1=z_2=\sqrt{\frac{m_1m_2}{2}}\frac{3\pi n l_P^3}{R_{11}} = x_5 = x_8, 
\ \ \ x_7=\frac{3\pi n l_P^3}{2 R_{11}}m_1
\eea
Clearly, the ellipsoid is very elongated in the $x_7$ direction, and in the 
supersymmetric case it degenerates into a line.

The second way to find the last term of the potential (\ref{v2}) is to 
consider an SO(3) invariant  
nonsupersymmetric configuration. The last term of (\ref{vt1}) represents a 
mass
given to 4 of the 5 scalars 
in our theory.  The most general scalar mass term can be written as:
\be
m^2( \Phi_5^2+\Phi_6^2+\Phi_8^2+\Phi_9^2+\Phi_7^2) + \mu_{ij} \Phi_i \Phi_j,
\label{general}
\ee
where the first piece is an $L=0$ combination and the second one is a 
combination
of various $L=2$ 
modes. In the supersymmetric case $\Phi_7$ is part of the same multiplet as 
the self-dual 3-form and half of 
the fermions, and remains massless, while the other 4 scalars receive 
identical masses. Thus, 
supersymmetry constraints the last term of (\ref{general}) to be 
\be
\mu_{ij} \Phi_i \Phi_j = {m^2 \over 4}(\Phi_5^2+\Phi_6^2+\Phi_8^2+\Phi_9^2- 
4 \Phi_7^2 )
\label{Lsusy}
\ee

In the nonsupersymmetric case only the $L=0$ mode is determined by the 
back reaction of the first order 
fields.  The $L=2$ modes can be specified independently in the boundary 
theory (this issue is discussed very briefly here, for a more 
thorough treatment we refer the 
reader to \cite{ps,ib2,ib3,zamora}).
Let us consider the case when both fermions have equal masses $m_1=m_2=m$.
We can turn off (\ref{Lsusy}) to restore the SO(5) symmetry between 
$ x_5, x_6, x_8, x_9$ and $x_7$, and turn on an $L=2$ mode which preserves the 
SO(3) symmetry in the polarization plane spanned by $ x_5, x_8, x_7$.
\be
- \Lambda(\Phi_5^2+ \Phi_8^2+ \Phi_7^2 - {3 \over 2} \Phi_6^2 -{3 \over 2} 
\Phi_9^2)
\ee
For $\Lambda$ of order $m^2$ or higher, a state in which the KKM is wrapped on 
a sphere of radius of 
order $r_0$ will have energy less than the vacuum energy. Thus, the M5 branes 
will polarize into a KKM 
wrapped on a 2-sphere of radius of order $r_0 \sim (n m l_P^3 )/R_{11} $, 
in the $578$ hyperplane.

When there is more than one shell, the potential is the sum of individual 
pieces of the form 
(\ref{v2}). One can also estimate the polarization potential for $Q$ 
coincident KKM shells and find that $r_{polarization} \sim r_0 /Q$. 
This configuration 
should also be part of the many vacua the perturbed theory should have.

Thus, the vacua of the theory will consist of multiple M5-KKM (ellipsoidal or 
spherical)  shells. We will concentrate in the rest of this paper on 
the SO(3) invariant nonsupersymmetric theories which 
correspond to polarization into a 
2-sphere. The generalization to the
almost supersymmetric theories can be easily done.

\subsection{The near shell solution}

$${}$$

The KK monopole solution without M5 charge, is given by the $R^7\times$ 
Taub-NUT space, 
\bea
ds^2&=&\eta_{\mu\nu}dx^{\mu}dx^{\nu}+V(r)(dx^{11}+4M(1-\cos \theta)d\phi)^2
\nonumber\\&+& 
 V(r)^{-1}(dr^2+ r^2 d\theta^2+r^2 \sin^2 \theta d\phi^2) \nonumber \\
V(r)^{-1}&=&1+ {4M \over r}. \label{oldm}
\eea
Near $r=0$ the metric $ds^2 \sim (dr^2 + r^2 d\theta^2  
+ r^2 \sin^2 \theta d\phi^2+ r^2 (d x^{11}/4M)^2)$ 
has a nut
singularity which is removed only if $x^{11}$ has period $16 \pi M = 
2 \pi R_{11}$. Thus the 
charge of the monopole is proportional to the length of the periodic isometry 
coordinate.
This explains why the effective tension of the monopole scales with 
$R_{11}^2$.
The 
requirement for the periodicity of $x^{11}$ can also be extracted from the 
requirement that there is no Dirac string singularity (this is explained in 
more detail in \cite{gross}).

One might be worried that upon adding a very large M5 charge to the monopole, 
the metric would be changed and the condition for not having a Dirac string 
would change from $R_{11} \sim M$, to some formula which also includes the M5 
charge. We thus need to examine the metric of a flat KKM with large M5 charge. 
This metric can be easily found \cite{papadopol} by lifting the 
D4-D6 metric \cite{myers2} obtained by T-dualizing smeared tilted D5 
branes. If we call $\alpha$ the original D5 tilting angle, and $\tilde N$ 
the original number of D5 branes, we obtain the flat KKM/M5 metric to be:
\bea
ds^2 =&& h^{-1/3} dx_{\parallel}^2 + h^{2/3}(dx_7^2+dx_8^2) + Z h^{-1/3}(dr^2+ 
r^2 d\theta^2+r^2 \sin^2 \theta d\phi^2 ) \nonumber \\
&&+h^{2/3}Z^{-1} (dx^{11}+\tilde N l_P\cos \alpha (1-\cos \theta) d \phi)^2, 
\label{newm}
\eea  
where 
\bea
Z&=&1+{\tilde N l_P\over r},\\ 
h&=&{1+\tilde N l_P/r \over 1 + \tilde N l_P\cos^2 \alpha /r }\label{zh}
\eea
 We can see that when $\cos \alpha =0$, $h$ and $Z$ are 
equal and we recover the metric of 
multiple M5 branes smeared on a 2-plane. When  $\cos \alpha = 1, h=1$ and 
we recover the plain KKM solution (\ref{oldm}). By comparing (\ref{newm}) with 
 (\ref{warp}) near the shell we can identify the 
coordinates $r$ and $\rho=\sqrt{(r_0-r_1)^2+r_2^2}$. We can also express 
the dummy coefficients $\tilde N$ and $\alpha$ in (\ref{zh}) in terms of our 
physical near shell  parameters by comparing (\ref{newm}) with (\ref{warp}) 
and (\ref{oldm}). We find the  near horizon limits of $Z$ and $h$:
\be
Z= {\pi l_p^3 N\over 2 r_0^2 \rho},\ \ \  \ \  
h= {\pi l_p^3 N\over 2 r_0^2 \left(\rho +{R_{11}^2 r_0^2 \over 2 
\pi l_p^3 N}\right)}
\label{zh2}
\ee
We can also express the factor $\  \tilde N \cos \alpha \ $ from (\ref{newm}) 
as:
\be
\tilde N l_P \cos \alpha = R_{11}/2.
\ee
Thus, $d x^{11}$ and $d \phi$ appear with the same relative 
coefficient as in in (\ref{oldm}) and therefore the condition for no 
Dirac string is unaffected by the presence of M5 charge. 

The KKM contribution to the metric becomes important only very close to the 
shell, at $\rho \sim {R_{11}^2 r_0^2 \over 2 \pi l_P^3 N} \sim  N m^2 l_P^3 
\equiv \rho_{crossover}$. We can 
see from (\ref{ro}) that ${\rho_{crossover}\over r_0}\sim m R_{11} \ll 1$, 
and thus the picture of the  KKM wrapped on a 2-sphere is consistent.

We can find the range of validity of the supergravity approximation by 
comparing $l_P$ to the radius of 
the transverse sphere in (\ref{newm}). We find supergravity valid if 
\be
\left({\rho\over \rho_{crossover}}\right)^{3/2}  N m^2 R_{11}^2 \gg 1,
\label{condr}
\ee
which is easily satisfied at large $N$, even for $\rho < \rho_{crossover} $. 
We also know that for  $\rho  < \rho_{crossover}$ the geometry becomes 
that of a KKM, which is regular everywhere. Thus, if supergravity is
valid at $\rho \sim \rho_{crossover}$, it is valid everywhere.

\section{The properties of the vacua of the perturbed theory}

$${}$$

We are now at the stage to begin exploring the properties of the M5-KKM vacua 
we found above. We 
know that the original (2,0) theory has little strings, 
and ``baryon vertices''. We expect to find 
these objects in the perturbed theory as well.
Moreover, since now we have many vacua we expect to have domain 
walls separating 
them. We would 
like to find the bulk M-theory configurations corresponding to these objects. 

\subsection{Little strings}

$${}$$

The nonperturbed (2,0) theory has two types of little strings. The
``quark'' little strings are massive, and correspond to M2 branes 
lowered from infinity and ending on the stack of M5 branes. The other
little strings are M2 branes going from one M5 brane to another, and are 
massless when the M5 branes are coincident.

The interaction energy between 
two ``quark'' little strings can be found by estimating 
the energy of the bulk M2 
brane whose boundary the strings are \cite{nea_ion}. In the UV, the M2 brane 
stays very near the AdS 
boundary, and thus does not know that the M5 branes sourcing the geometry are 
polarized. This is 
expected - giving a small mass to massless fields should not affect UV physics.
In the infrared, the M2 brane gets  near the M5-KKM shell and interacts with 
it, and thus the physics will change. 

Let us consider two largely separated parallel strings which are the boundary 
of an M2 brane which
comes very near the KKM-M5 shell. In the limit when the strings are very far 
apart, the largest 
contribution to the interaction energy comes from the part of the M2 brane 
which is parallel to the shell. 

Since we have broken Lorentz invariance by turning on an R-current, the 
interaction energy will depend 
on the orientation as well as on the direction of separation of the little 
strings.  

For the sake of clarity, let us explain in a bit more detail what the setup is.
The polarized M5's span the $0,1,2,3,4, 11$ directions, with
the 11th direction compactified on a circle of radius $R_{11}$. The KKM
in which the M5's polarize has the topology $R^4\times S_2$, with the 
11th direction being the isometry direction. If we use polar coordinates in 
the plane of the $S_2$, the KKM is aligned along $0,1,2,3,4, \theta,\phi$. 
The geometry at infinity is still $AdS_7\times S_4$,
with the boundary of $AdS_7$ spanned by  $0,1,2,3,4,11$. The radial 
coordinate of $AdS_7$ is $r$. 

We distinguish two possible situations, depending on  the orientation and 
separation of the ``quark'' little strings:

i) None of the directions of the M2 brane is 11. This corresponds to  non-11 
strings separated along 
any direction but 11. The action of the M2 
brane is proportional to the square 
root of the determinant of the induced metric. From (\ref{newm})  we find this 
to be 
$\sqrt{\det(G_{\rm{induced}})} = h^{-1/2}$. Since $h$ remains finite as we 
approach the shell 
($\rho\rightarrow 0$), the interaction energy of two little strings is 
proportional
to their separation. Thus the 
strings are confined.

One can easily estimate the tension of the confining ``flux tube'', 
 However, there is a caveat - the supergravity approximation is valid at  
 $\rho \sim \rho_{crossover}$ only for 
$N m^2 R_{11}$ large enough (\ref{condr}).
 Unless this condition is satisfied, we cannot use  
 (\ref{newm}) at $r \approx \rho_{crossover}$ to  reliably 
compute this tension. 
Thus, 
\be
T=T_{M2} h^{-1/2}|_{min} \sim {r_0^2 R_{11} \over l_p^6 N} \sim {m^2 N\over 
R_{11}}.
\ee

ii) One of the direction of the  M2 brane is 11. This corresponds to 
either non-11 strings separated 
along the 11'th directions, or to (0,11) strings separated along any other 
direction. In this case
$\det(G_{\rm{induced}})=Z^{-1/2}$.
Since  $Z^{-1}$ goes to zero as $r\rightarrow 0$, it does not cost any energy 
to take the strings further 
apart. Thus the strings are ``screened''. The M-theory explanation of 
this phenomenon is that the 2 sides 
of the  M2 brane attach to the shell, and they are free to move apart.

We also expect the tensionless strings to get an effective tension as 
we turn on $m$. We can use a naive 
argument to estimate this tension. The original tensionless strings come 
from M2 branes stretched 
between coincident M5 branes. When the $N$ M5 branes are spread on a 
2-sphere of radius $r_0$, the 
average separation between them will be of order $r_0 N^{-1/2}$, 
which will give an effective tension of order
\be
T_{effective} \sim r_0 N^{-1/2} (G_{55}G_{00}G_{ii})^{1/2}=r_0 N^{-1/2}
\ee
for (0,i) little strings, and
\be
T_{effective} \sim r_0 N^{-1/2} (G_{55}G_{00}G_{11,11})^{1/2}=r_0 N^{-1/2} h 
Z^{-1} 
\ee
for (0,11) little strings. It appears that the (0,11) strings remain massless.
However, more is happening. At the KKM core the size of the 11'th dimension
is zero, and thus a string originally winding on the 11'th direction disappears
from the spectrum.

\subsection{Domain walls}

$${}$$

The perturbed theory has many isolated vacua, which must 
be separated by domain walls. If the two vacua have the same number of KKM's 
but different distributions of M5 charge, the KKM's  will generically 
intersect on a 2 sphere on which they will exchange M5 charge. 
This bulk configuration is dual to the boundary theory domain wall.  
The tension of the domain wall
is given by the bending tension of the configuration.

If the vacua have different numbers of KKM's, $Q_1\neq Q_2$, the monopoles 
intersect again on a 2-sphere where they exchange M5 charge. By KKM charge 
conservation, a KKM with charge $Q_2-Q_1$ will fill the 3-ball whose 
boundary the 2-sphere is. The tension will have an extra piece coming 
from the energy of the KKM's which fill the 2-sphere.

If we do not have any supersymmetry, the two vacua separated by the domain wall
will have generically different energies, and thus the domain wall will be 
moving. In the limit
in which supersymmetry is restored, the energies of the vacua are almost the 
same, and the domain
walls will be almost stationary

\subsection{Baryon vertices}

$${}$$

In the unperturbed $AdS_7 \times S^4$ background
the string baryon vertex is an M5 brane
wrapped on $S_4$, with $N$ M2 branes lowered from the
boundary ending on it. This configuration has been discussed thoroughly in 
\cite{ali,town}.

Let us analyze what happens to the baryon vertex
in the background of polarized M5's. In the UV the physics is not affected 
by the
perturbation, and thus the baryon vertex corresponds to an M5 brane wrapped 
on $S_4$. 
As we flow to the IR, the wrapped M5 eventually reaches the KKM shell and 
crosses it. In order to see if any objects are created via the Hanany-Witten 
effect \cite{hw}, let us ``zoom in'' on a region near the shell where the 
branes look almost flat and $\hat r \parallel \hat x_5$.

Since the 11'th direction was made special by turning on an R-current, 
we will first analyze a baryon vertex which is not extended along the 
11'th direction, and then analyze one which is.
Let us consider a baryon vertex M5 brane locally extended along the 
$0,1,6,7,8,9$
directions, and a KKM locally extended along  $0,1,2,3,4,6,7$. We claim that 
when we 
bring the M5 brane along $x_5$ and cross the KKM another M5 brane is created, 
extended in the $0,1,5,6,7,11$ directions. 
It appears puzzling that the new brane is extended along the 11'th direction. 
Nevertheless, we believe that this is indeed the case. 
There are three arguments one can give to support this. 
The first is that one obtains this configuration by lifting to 11 dimensions 
the 
Hanany-Witten effect at the crossing of an NS5-brane and a D6-brane, when
a D4-brane is created transverse to the other two.
The second is that, the KKM  end of the created brane is  magnetic 
with respect to the 2-form field strength on the monopole, as it should be. 
The third one will follow when we investigate the properties of the new 
baryon vertex.

As one brings the wrapped M5 through the KKM shell, the M2 branes hanging from 
the boundary will end on the intersection of the created M5 with the KKM, 
and there will be nothing preventing the original wrapped M5 to shrink to 
zero size.  

Thus, the bulk dual of the new baryon vertex is an M5 brane which fills the
3-ball whose boundary the KKM 2-sphere is, and which has one direction along 
the
isometry direction. To justify the baryon vertex name, we look at the KKM 
action
(\ref{k7}) term
\be
\int d^7 \xi Dx^\Lambda\we Dx^\Pi\we Dx^\Sigma\we 
C_{\Lambda\Pi\Sigma}\we db\we db
\ee
where $b$ is the one-form on the worldvolume of the monopole.
Let us denote $db$ by $F_2$. Due to the magnetic ending of the created M5, 
the Bianchi identity gets modified and  this term will no
longer be gauge invariant under the background gauge transformation 
$\delta C_3=d\chi_2$:
\be
\delta\int F_2(\rm{monopole})\we F_2(\rm{dissolved})\we C_3=-
\int d F_2(\rm{monopole})\we F_2(\rm{dissolved})\we \chi_2
\ee
unless $N$ M2's descend from the boundary to the M5-KKM junction.

One might also be worried that the new baryon vertex is extended along 
three directions (0,1,11), and might have a higher energy than the original 
baryon vertex which is only extended along (0,1). We can calculate this 
energy  using the induced metric (\ref{warp})
\bea
&&E=\frac{1}{l_P^6}R_{11} 4\pi\int dx^0 dx^1 dr_1 r_1^2
Z^{1/2}\nonumber\\
&&=\frac{4\pi R_{11}}{l_P^6}\int dx^0  dx^1 \int_0^{r_0} dr_1 r_1^2
\sqrt{\frac{R^3}{2r_0 r_1}\left(\frac{1}{r_0-r_1}-\frac{1}{r_0+r_1}\right)}
\nonumber\\
&&=\int dx^0 dx^1 \frac{\sqrt 2 \pi^2 R_{11}(R r_0)^{3/2}}{l_P^6}\label{enew}
\eea
Substituting $r_0$ we obtain the new baryon vertex energy:
\be
E\sim \int dx^0 dx^1
\sqrt{N } m^{3/2} R_{11}^{-1/2}
\ee
The energy of an M5 brane wrapped on $S^4$ of radius $\rho_{crossover}$, 
namely
the old baryon vertex at the point where  M-theory phenomena like the 
Hanany-Witten effect described above appear, is 
\be
E\sim T_{M5}\int dx^0 dx^1 Z(\rho) \rho_{crossover}^4\sim
\int dx^0 dx^1 N^2 m^4 R_{11}^2  
\ee
where we used the near shell metric to compute the induced metric on
$S^4$.

Thus the condition that the new configuration 
has a lower energy is
\be
1\ll N^{3/2} (mR_{11})^{5/2}
\ee
which is easily satisfied for sufficiently large $N$.

When the UV baryon vertex is extended along $x_{11}$, there does not seem 
to be any 
object created when it passes through the KKM shell. This is consistent with 
the fact that strings extended along $x_{11}$ are screened.  Once 
the baryon vertex gets near the KKM shell these strings can attach to the it, 
and there is nothing to prevent the M5 brane from collapsing. Thus the 
baryon vertex extended along $x_{11}$ disappears in the IR.


\section{Conclusions and future directions}

$${}$$

We investigated the effect of a fermion R-current perturbation on 
the (2,0) theory living on the worldvolume of $N$ coincident M5 branes. 
We found the dual M-theory background to contain the M5-branes 
polarized 
into Kaluza-Klein monopoles, with the isometry direction of the monopoles
on the worldvolume of the M5-branes.
However, we showed that polarization is possible only when all the
supersymmetry is broken.

 We examined the properties of 
the vacua of the perturbed theory. We discovered that (0,i) tensionless 
little strings acquire a tension, while the (0,11) tensionless little strings 
disappear from the spectrum.

The ``quark'' little strings extended along the isometry directions are 
screened, and the strings which are not extended 
along
$x^{11}$ are confined. We also found the tension of the confining flux tube

We also found the new IR description of the string baryon vertex, by 
flowing the UV baryon vertex (which is an M5-brane 
wrapped on $S_4$) to the IR. A new M5-brane is formed when the 
UV vertex crosses the M5-KKM shell. This new 
baryon vertex fills the 3-ball whose boundary the KKM is, 
and is extended along the isometry direction. The remaining confined little
strings will have precisely this baryon vertex. 
We checked that the new baryon configuration is energetically 
preferred to the old one. 
We have also given an M-theory description to the 
domain walls in the perturbed theory.

This paper completes a missing piece in the understanding of  brane 
polarization, by showing that M5 branes can and do polarize into Kaluza-Klein
monopoles. The new configuration might be a good setup  for an entropy 
computation in the spirit of 
\cite{freedman} to see if the correction to the entropy has $N^3$ behavior.

It is also interesting to ask if the M5 $\ra$ KKM polarization exhausts 
the series of possible 
brane polarizations. Besides the Dp $\ra$ D(p+2), the D2,D3 $\ra$ NS5 
polarizations explored in \cite{ps,ib2}, and the F1 $\ra$ D4 polarization 
which
is the reduction of the M2 $\ra$ M5 polarization studied in \cite{ib1}, 
it looks like type IIA F1 strings and D4 branes might polarize into D2 branes 
and
respectively NS5 branes, wrapped on a one-sphere. This is supported by the fact
that a D2/NS5 brane wrapped on a one-sphere can have quantized F1/D4 charge. 
These possible polarizations would be however different from the ones studied 
so far, for two reasons.

First, the F1 string and the D2 brane come from the same object in M-theory. 
Lifting a D2 wrapped on a one sphere with a large F1 charge to 11 dimensions 
one obtains a helix shaped M2 brane wrapped on the 11'th direction. 
Even if this configuration might be stable in a perturbed theory, it is quite 
unlikely that it will be an isolated vacuum like before.

Second, one can also find the powers of $r$ which appear in the polarization 
potential (the equivalent of \ref{vt1}) \footnote{We thank W. Taylor for 
bringing
this to our attention.}. The potential will contain three $r^2$ terms. 
Thus, in the supersymmetric case this potential will either vanish (in which 
case
we expect to have a moduli space of possible polarization vacua) or have only 
one minimum at $r=0$ (which means no polarization). In the nonsupersymmetric 
case
a static configuration might exist at a nonzero radius (determined by the 
balance
of potential terms from other sources). We believe it is worth investigating 
whether these  polarization states exist at all, and what are their properties.

{\bf Acknowledgments:} We are very grateful to Joe Polchinski for 
numerous discussions and reading the manuscript. We would also like
to acknowledge useful conversations with  Aleksey Nudelman, 
Alex Buchel, Wati Taylor, and Radu Roiban. The work of I.B. was 
supported in part by NSF grant PHY97-22022 and the work of D.V. 
by NSF grant 9722101.

\section{Appendix}

We list several properties of the antisymmetric 2- and 3-tensors 
which form a basis for forms on the transverse space.

\bea
V_3&=&\frac{1}{3!}(\frac{x^q x^m}{r^2} T_{qnp}+{\rm{2\; more}})
dx^m\we dx^n \we dx^p\\
V_2&=&\frac{1}{2!}(\frac{x^qx^i}{r^2}T_{qj}+{\rm{1 \;more}})
dx^i\we dx^j\\
T_2-V_2&=&*_5 V_3\\
d(\ln r)\we V_3&=&0\\
d(\ln r)\we *_5 V_3&=&d(\ln r)\we *_5 T_3\\
d(V_3)&=&-3d(\ln r)\we T_3\\
d(*_5 V_3)&=&2 \ln r\we *_5 T_3
\eea


\end{document}